\documentstyle[prl,aps,epsf,multicol]{revtex}
\begin{document}

%
\newcommand{\fig}[2]{\epsfxsize=#1\epsfbox{#2}}
%
%
%
\newcommand{\passage}{
\end{multicols}\widetext\noindent\rule{8.8cm}{.1mm}%
  \rule{.1mm}{.4cm}} 
 \newcommand{\retour}{
\noindent\rule{9.1cm}{0mm}\rule{.1mm}{.4cm}\rule[.4cm]{8.8cm}{.1mm}%
         \begin{multicols}{2} }
 \newcommand{\unecol}{\end{multicols}}
 \newcommand{\deuxcol}{\begin{multicols}{2}}

\tolerance 2000

\author{Pascal Chauve{$^1$}, Pierre Le Doussal{$^2$} and Kay Wiese{$^3$}}
\address{{$^1$} CNRS-Laboratoire de Physique des Solides, Universit{\'e} de
Paris-Sud, B{\^a}t. 510 , 91405 Orsay France}
\address{{$^2$}CNRS-Laboratoire de Physique Th{\'e}orique de 
l'Ecole Normale Sup{\'e}rieure,
24 rue Lhomond,75231 Cedex 05, Paris France.}
\address{{$^3$}FB Physik, Universi\"at Essen, 45117 Essen, Germany}

\title{Renormalization of pinned elastic systems: how does it work beyond
one loop ?}
\date{\today}
\maketitle

\begin{abstract}
We study the field theories for pinned elastic systems at equilibrium and at depinning.
Their $\beta$-functions differ to two loops by novel ``anomalous'' terms. At equilibrium we find a roughness
$\zeta=0.20829804 \epsilon + 0.006858 \epsilon^2$ (random bond),
$\zeta=\epsilon/3$ (random field). At depinning we prove two-loop renormalizability 
and that random field attracts shorter range disorder. We find
$\zeta=\frac{\epsilon}{3}(1 + 0.14331 \epsilon)$, $\epsilon=4-d$, in violation of
the conjecture $\zeta=\epsilon/3$, solving the discrepancy with simulations. 
For long range elasticity $\zeta=\frac{\epsilon}{3}(1 + 0.39735 \epsilon)$, $\epsilon=2-d$,
much closer to the experimental value ($\approx 0.5$ both for liquid helium contact line depinning
and slow crack fronts) than the standard prediction $1/3$.
\end{abstract}

\deuxcol

The aim of this Letter is to report progress on a conceptual issue
and, as a byproduct, to resolve a longstanding discrepancy between theory and
numerical simulations or experiments.
The issue is whether it is possible to construct a field theory of disordered elastic systems,
at equilibrium and at depinning, renormalizable beyond one loop
as for standard critical phenomena. A discrepancy exists at present 
between the value for the roughness exponent $\zeta$ predicted by
theory ($\zeta=\epsilon/3$ exactly) and simulations as well as experiments 
on wetting and on cracks.

Numerous experimental systems can indeed be modelled as elastic objects pinned
by random impurities, with specific features. 
Interfaces in magnets  \cite{nattermann_book_young} 
experience either random bond RB (i.e. short range) disorder or
random field RF (i.e. long range) disorder. Charge density waves (CDW) or the Bragg glass
in superconductors \cite{tgpldbragg} are periodic objects. The contact line of liquid helium
meniscus on a rough substrate is governed by long range elasticity and so are slowly
propagating cracks \cite{rolley,ertas_kardar,fishercrack,schmittbuhl}. They can all be parametrized 
by a height (or displacement) field $u(x)$ 
($x$ being the $d$-dimensional internal coordinate of the elastic object),
with in some cases $N>1$ components. The roughness exponent $\zeta$:
\begin{eqnarray}
|u(x) - u(x')| ^2 \sim |x-x'|^{2 \zeta}
\end{eqnarray} 
is measured in experiments for systems 
at equilibrium ($\zeta_{\rm eq}$) or driven by a force $f$.
Other exponents describe the velocity near the depinning threshold $f_c$, $v \sim (f-f_c)^\beta$, the scaling
of the dynamical response, $t \sim x^z$, and the local velocity 
correlation length $\xi \sim (f-f_c)^{-\nu}$.

The study of pinned elastic systems, among a broader class of disordered models 
(e.g. random field spin models), is notably difficult due to dimensional
reduction DR which renders naive perturbation theory useless \cite{nattermann_book_young,dimred2}. Indeed
to {\em any} order in the disorder at zero temperature $T=0$, any physical observable 
is found to be {\it identical} to its (trivial) average in a Gaussian random 
force (Larkin) model. A bold way out of this puzzle was proposed by Fisher \cite{fisher_functional_rg} 
within a one-loop renormalization group analysis of the interface problem in $d=4 -\epsilon$.
He noted that the coarse grained disorder correlator becomes {\em non-analytic}
beyond the Larkin scale $L_c$, yielding large scale results distinct from 
naive perturbation theory. An infinite set of operators 
become relevant in $d<4$, parameterized by 
the second cumulant $R(u)$ of the random potential, 
i.e.\ $\overline{V(x,u) V(x',u')} = \delta_{x-x'} R(u-u')$.
Explicit solution of the one-loop Functional RG equation (FRG) for $R(u)$
gives several non trivial attractive fixed points (FP)
to ${\cal O}(\epsilon)$ proposed in 
\cite{fisher_functional_rg} to describe RB, RF disorder and in \cite{tgpldbragg},
periodic systems (RP) such as CDW or vortex lattices. All these FP exhibit a ``cusp'' singularity
as $R^{*\prime \prime}(u) - R^{*\prime \prime}(0) \sim |u|$ at small $|u|$. 
Large $N$ and variational
methods \cite{mezard_parisi,tgpldbragg}
confirmed the picture and the cusp was
interpreted in terms of shocks in the renormalized force \cite{balents_rsb_frg}.
A FRG was also developed to 
one loop \cite{leschhorn_depinning2,narayan_fisher} to 
describe the {\it driven dynamics}  just above 
depinning $f=f_c^+$, the cusp being linked to the threshold $f_c \sim |\Delta'(0^+)|$.
Surprisingly, the flow equation for the correlator $\Delta(u)$ 
of the force $F(x,u)$ is, to one loop, {\em identical} to the 
one of the statics (with $\Delta(u) = - R''(u)$). Extension to temperature
$T>0$ yielded rounding of the cusp in a layer $u \sim T$
and the celebrated creep law \cite{chauve_creep}.

Despite these successes, serious difficulties remain. 
First, in the last fifteen years since \cite{fisher_functional_rg},
no study has addressed whether the FRG yields, beyond one loop, a
renormalizable field theory able to predict universal results
\cite{footnote_larkin}. Doubts were even raised \cite{balents_fisher} about the
validity of the $\epsilon$-expansion beyond the order ${\cal O}(\epsilon)$.
Second, numerous simulations near 
depinning \cite{leschhorn_depinning2,leschhorn,roters,nowak_usadel_2d} seem to exclude
$\zeta=\epsilon/3$ argued in \cite{narayan_fisher} to be exact. In the case of long range elasticity, the 
prediction $\zeta=(2-d)/3$ \cite{ertas_kardar}  {\it disagrees} 
with the systematically larger value $\zeta \approx 0.55$ ($d=1$) measured 
for liquid Helium contact line depinning \cite{rolley}
and for the in plane roughness of slow crack fronts \cite{schmittbuhl} (see also
simulations \cite{paczuski}).

In this Letter, we address these issues both for dynamics and statics.
The main difficulty is the non-analytic nature of the theory (i.e. the
fixed point action) at $T=0$, which makes it a priori quite different 
from conventional critical phenomena. For {\it depinning}, we overcome
the problem and show renormalizability at two-loop order. As a result
we resolve several questions left unclear in previous works. We find
that (i) quasi-static driven dynamics differs from statics at two loops
(ii) shorter range disorder is within the RF universality class 
and (iii) the conjecture $\zeta=\epsilon/3$ is violated. This last result
resolves the longstanding discrepancy with simulations. In the case of
long range elasticity it yields $\zeta \approx 0.5$ for $d=1$ and may
thus explain the high value of $\zeta$ found in experiments on
cracks and wetting. For the {\it statics} we find ambiguities at $T=0$ 
which we lift using a renormalizability
condition, yielding fixed points and $\zeta_{\rm eq}$ to ${\cal O}(\epsilon^2)$. This
result is also obtained within an independent exact FRG study
\cite{chauve_pld}. The FRG equation for the disorder contains new 
anomalous terms both for statics and dynamics, which are absent 
in an analytic theory. Our predictions for all exponents are shown in Tables I,II.

\begin{center}

\noindent
\begin{tabular}{|c|c|c|c|c|r|}
\hline 
 & $d$ & $\epsilon$ & $\epsilon^2$ & estimate & simulation~~~\\
\hline 
\hline 
        & $3$ & $0.33$ & 0.38 & 0.38$\pm$0.02 & 0.34$\pm$0.01 \cite{leschhorn_depinning2} \\
\hline 
$\zeta$ & $2$ & 0.67 & 0.86 & 0.82$\pm$0.1 & 0.75$\pm$0.02 \cite{leschhorn} \\
\hline 
        & $1$ & 1.00 & 1.43 & 1.2$\pm$0.2 & 1.25$\pm$0.05 \cite{leschhorn}  \\
\hline 
\hline 
        & $3$ & 0.89 & 0.85 & 0.84$\pm$0.01 & 0.84$\pm$0.02 \cite{leschhorn_depinning2} \\
\hline 
$\beta$ & $2$ & 0.78 & 0.62 & 0.53$\pm$0.15  & 0.64$\pm$0.02 \cite{leschhorn_depinning2}
\\
\hline 
        & $1$ & 0.67 & 0.31 & 0.2$\pm$0.2 & $\approx$ 0.3 \cite{leschhorn,nowak_usadel_2d} \\
\hline 
\hline 
        & $3$ & 0.58 & 0.61 &   0.62$\pm$0.01  & \\
\hline 
$\nu$ & $2$ & 0.67 & 0.77 &  0.85$\pm$0.1   & 0.77$\pm$0.04 \cite{roters} \\ 
\hline 
        & $1$ & 0.75 & 0.98 &   1.25$\pm$0.3  & 1$\pm$0.05 \cite{nowak_usadel_2d}  \\ 
\hline
\hline
\hline
& $3$  & 0.208 &  0.215  & $0.215\pm 0.003$    & $0.22\pm 0.01$ \cite{middleton}  \\
\hline
$\zeta _{\rm eq}$ & $2$ &0.417 &0.444 & $0.438\pm 0.007$ & $0.41\pm 0.01$ \cite{middleton} \\
\hline
& $1$ & 0.625 & 0.687 & 2/3 & 2/3  \\
\hline
\end{tabular}

\end{center}

\vspace{1mm}

\noindent
{\small {\bf Table I}: exponents for depinning and statics ($\zeta_{\rm eq}$)
as obtained, respectively: from setting $\epsilon=4-d$ in the one loop and 
two loop result, from Pad{\'e} estimates together with scaling relations
and from numerical works. For $\zeta _{\rm eq}$ we have improved the
estimate using the exact result $\zeta _{\rm eq}(d=1)=2/3$.}

\vspace{2mm}

\begin{center}

\noindent
\begin{tabular}{||c|c|c|c||c|c|c|c||}
\hline
 & $\epsilon$ & $\epsilon^2$ & estimate  & & $\epsilon$ & $\epsilon^2$ & estimate \\
\hline
$\zeta $  & 0.33     &  0.47    & $0.5\pm 0.1$  & $\beta $ & 0.78 & 0.59 & $0.4\pm 0.2$  \\
\hline
$z$ &0.78 &0.66 &$0.7\pm 0.1$ & $\nu $ & 1.33 & 1.58 & $2.\pm 0.4$ \\
\hline
\end{tabular}

\end{center}

\vspace{1mm}

\noindent
{\small {\bf Table II}: depinning exponents for long range elasticity in $d=1$:
$\zeta$ is consistent with experiments on contact line depinning ($\zeta \approx 0.5$ \cite{rolley})
and cracks ($\zeta \approx 0.55 \pm 0.05$ \cite{schmittbuhl}). }

The starting point is the equation of motion:
\begin{eqnarray}
\eta \partial_t u_{xt} = \partial_x^2 u_{xt}  + F(x,u_{xt} ) 
\end{eqnarray}
with friction $\eta$
and in the case of long range elasticity we replace (in Fourier) 
$q^2 u_q$ by $|q| u_q$ in the elastic force.  Disorder averaged correlations 
$\overline{\langle A[u_{xt}]\rangle}= \langle A[u_{xt}]\rangle_S$ 
and responses $\overline{\delta \langle A[u]\rangle/\delta h_{xt}}= \langle 
\hat{u}_{xt} A[u] \rangle _{S}$ can be computed from the standard averaged dynamical action:
\begin{eqnarray}
&& S= \int_{xt} \hat{u}_{xt} (\eta \partial_t 
- \partial_x^2) u_{xt} - \frac{1}{2} \int_{xtt'}
\hat{u}_{xt} \hat{u}_{xt'} \Delta(u_{xt}- u_{xt'})  \nonumber 
\end{eqnarray}
Finite temperature is studied adding $- \eta T \int_{xt} \hat{u}^2_{xt}$,
driven dynamics adding $- f \int_{xt} \hat{u}_{xt}$ and shifting $u \to u + v t$ 
in $S$.  We study the quasi-static limit $v=0^+$,
as well as equilibrium dynamics $f=0$ where, via fluctuation dissipation relations, static
quantities can be equivalently computed using $S$ or
the replicated hamiltonian \cite{footnote4}.

It is useful to first study naive perturbation theory,
in  an {\em analytic} $\Delta(u)$ i.e. in its
derivatives $\Delta^{(n)}(0)$, using 
the diagrammatic rules of Fig.\ \ref{fig1}. Since at 
each vertex there are one conservation rule for momentum and two for frequency
we consider both unsplitted (local $x$) and
splitted (bilocal $t,t'$) vertices (and splitted $a,b$ vertices
in the statics). $T=0$ power counting yields 
$\int_t \hat{u} u \sim x^{d-2}$ and $u \sim x^{\zeta}$, where 
$\zeta= {\cal O}(\epsilon=4-d)$ has to be determined.
For an analytic $\Delta(u)$ the perturbation expansion 
of any (analytic) observable yields identical results
\cite{footnote6} as setting $\Delta(u) \equiv \Delta(0)$
and one obtains the incorrect DR roughness $\zeta=\epsilon/2$. Temperature is formally irrelevant and must be scaled
\cite{footnote7} as $T = \tilde{T} \Lambda^{-2+\epsilon-2 \zeta}$
with the UV cutoff $\Lambda$ (and fixed dimensionless $\tilde{T}$).
By power counting the only superficially UV divergent  irreducible vertex 
functions (IVF) are found to involve only one or two response fields
$\hat{u}$ (at $T>0$ each $\tilde{T}$ comes with a required
$\Lambda^{2-d}$ factor to compensate the 
divergence \cite{footnote7}).
The statistical tilt symmetry $u_{xt}\to u_{xt}+\mbox{const.}$
(see e.g.\ \cite{leschhorn_depinning2,narayan_fisher})
further restricts the needed counterterms 
at $f=f_c$ to only one for $\eta$ and one for the full function $\Delta(u)$.
The one loop (D)
and two loops (A,B,C) diagrams which correct the disorder at $T=0$ are shown in 
Fig.\ \ref{fig1} (unsplitted).
The splitted graphs corresponding to A in the statics (and
which do not vanish or cancel in what follows)
are shown in Fig.\ \ref{fig2}.
The dynamical diagrams are obtained from the
static ones by adding one external $\hat{u}$ on each connected component
(e.g.\ $b$ generates $b_1,\ldots,b_6$).
To escape triviality at $T=0$ we must now develop perturbation theory in
a non-analytic interaction $\Delta(u)$ (or $R(u)$),
a non trivial extension of conventional field theory.
Let us illustrate the new rules. 
Derivation 
\begin{figure}[htb]
\centerline{ \fig{6cm}{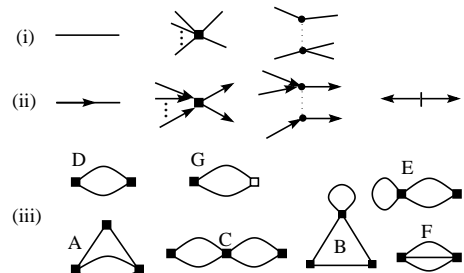} }
\caption{{\narrowtext (i) diagrammatic
rules for the statics: replica propagator $\langle u_a
u_b\rangle_0 \equiv T\delta_{ab}/q^{2}$, 
unsplitted vertex, equivalent splitted vertex $-\sum_{ab} \frac{1}{2
T^2} R(u_a-u_b)$. (ii) dynamics: response propagator $\langle \hat{u} u
\rangle_0 \equiv R_{q,t-t'}$,
unsplitted vertex, splitted vertex $-\frac{1}{2} \hat{u}_{xt} \hat{u}_{xt'}
\Delta(u_{xt}-u_{xt'})$ 
and temperature vertex. Arrows are along increasing time. An 
arbitrary number of lines can enter these functional
vertices. 
(iii) unsplitted diagrams to one loop D, with inserted 
counterterm G and two loop A,B,C,E,F.}}
\label{fig1}
\end{figure}%
\begin{figure}[htb]
\centerline{\fig{8cm}{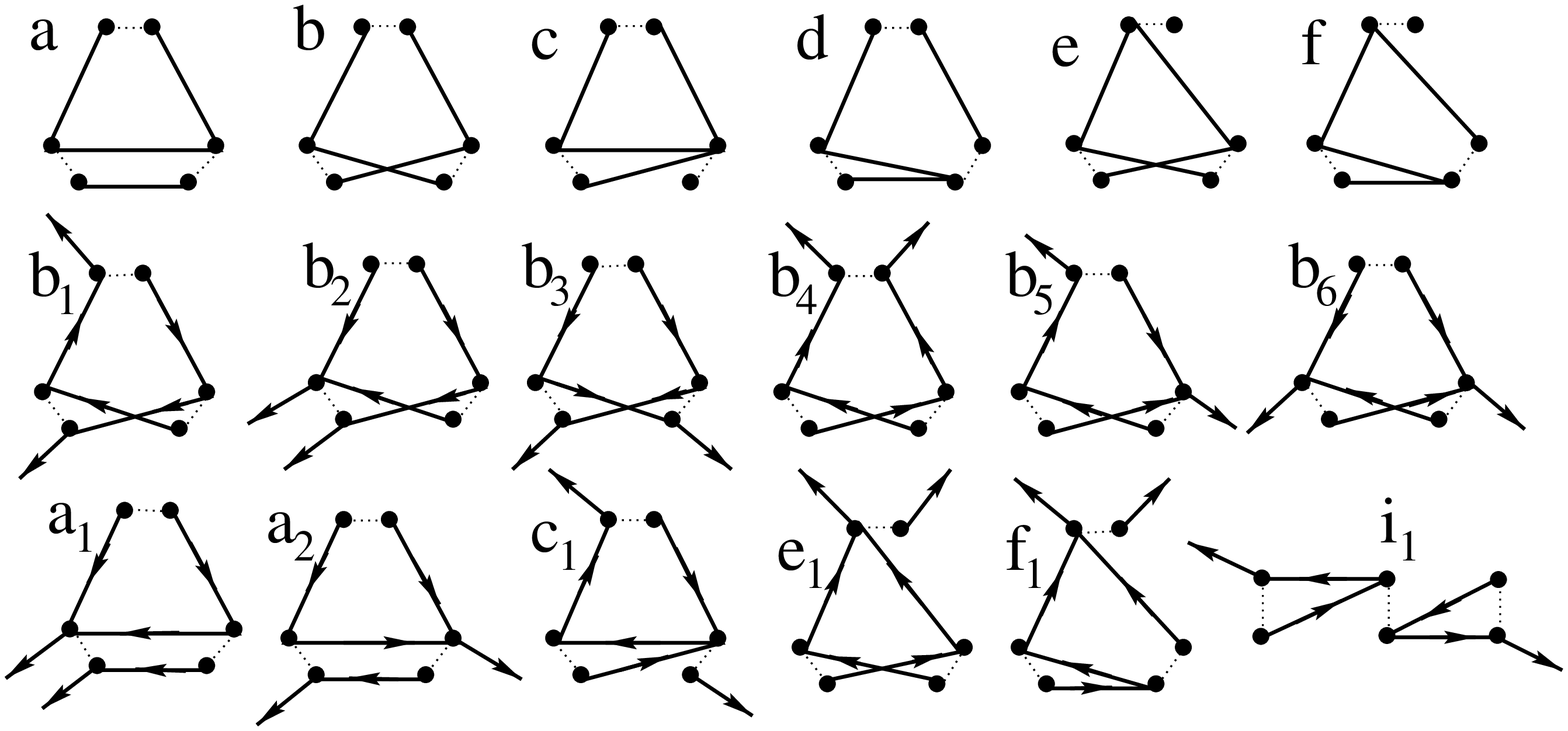}}
\caption{(a-f): the six splitted (static) diagrams corresponding to
two loop A  
diagram. Below: the corresponding non vanishing diagrams in the
dynamics. The last one is 
the only non trivial C diagram (see text). \label{larkinhatstat}}
\label{fig2}
\end{figure}%
\noindent
by extracting a leg from a vertex can be done as  
usual only for a vertex 
evaluated at a generic $u$ (e.g.\ graphs $b_i$ in Fig~\ref{fig2}).
If it is evaluated at $u=0$ (e.g.\ graph $e_1$), one must expand
$\Delta(u)$ in powers of $|u|$, i.e.\ $\Delta(u) = \Delta(0) +
\Delta'(0^+) |u| +  \Delta''(0^+) u^2 /2 + \ldots$ and 
carefully apply Wick's rules. The result is
that the above diagrammatic rules (Fig.\ \ref{fig1},\ref{fig2}) can still be used  except that 
the values of the diagrams are {\em different}. 
The graphs of Fig.~\ref{fig2} correspond to performing four Wick 
contractions and some 
end up in evaluating non trivial averages of e.g.\ sign or delta functions.
For instance $e_1$, which vanishes in the analytic
theory since $\Delta'(0)=0$, now reads: 
\begin{eqnarray}
&& e_1 = \Delta'(0^+)^2 \Delta''(u) \int_{t_i>0,r_i}
\!\!\!\!\!\!\!\!\!\!\!\!\!
R_{r_1,t_1} R_{r_1,t_2} R_{r_3-r_1,t_3} R_{r_3,t_4} F_{r_i,t_i} \nonumber
\ ,
\end{eqnarray}
where $F_{r_i,t_i}=\langle {\rm sgn}(X) {\rm sgn}(Y)\rangle$, $X =
u_{r_1,-t_3}-u_{r_1,-t_4-t_1}$, 
$Y = u_{0,-t_4}-u_{0,-t_3-t_2}$, computed with Gaussian
averages. The limit $T \to 0$ at $v=0$ yields
$\langle {\rm sgn}(X) {\rm sgn}(Y) \rangle=\frac{2}{\pi}
{\rm asin}(\langle XY \rangle/\sqrt{\langle X^2\rangle \langle Y^2 \rangle}))$,
and a complicated $T=0$ expression for
$e_{1}$ in the statics \cite{us_long}. The opposite limit $v \to 0$ at $T=0$
corresponds to depinning, with 
$\langle {\rm sgn}(X) {\rm sgn}(Y) \rangle  \to {\rm sgn}(t_4+t_1-t_3)
{\rm sgn}(t_3+t_2-t_4)$, and more generally to 
$\Delta^{(n)}(u_t - u_{t'}) \to \Delta^{(n)}(v(t - t'))$ in any
vertex evaluated at $u=0$.

We now focus on depinning at $T=0$.
Using these rules we compute in
perturbation of $\Delta \equiv \Delta(u)$ the contributions to the disorder IVF 
to one and two loops:
\begin{eqnarray}
&& \delta^{1} \Delta = - (\Delta'^2 + (\Delta - \Delta(0)) \Delta'' ) I \\
&& \delta^{2} \Delta = ((\Delta - \Delta(0)) \Delta'^2 )'' I_A  \label{d21} \\
&& + \frac{1}{2} ((\Delta - \Delta(0))^2 \Delta'' )'' I^2
+ \Delta'(0^+)^2 \Delta'' (I_A - I^2)  \label{d22}
\end{eqnarray}
with $I=\int_q 1/q^4$ and $I_A = \int_{q_1,q_2} 1/q_1^2 q_2^4 (q_1+q_2)^2$
\cite{footnote17},
whose divergent parts $\delta_{div}^{1} \Delta$, 
$\delta_{div}^{2} \Delta$ yield the one loop
and two loop counterterms respectively. These 
are computed here adding a mass $q^2 \to q^2 + m^2$, using dimensional regularization
$I m^\epsilon= N_d (\frac{1}{\epsilon} 
+ {\cal O}(\epsilon))$,
$I_A m^{2\epsilon} = N_d( \frac{1}{2 \epsilon^2} 
+ \frac{1}{4 \epsilon})$ and absorbing $N_d=(d-2)/(4 \pi)^{d/2}\Gamma(\frac{d}{2})$
in $\Delta$.
(\ref{d21}) comes from $a_1+a_2+\sum_i b_i$, the first term in  
(\ref{d22}) from all graphs C (not detailed) {\em except} graph
$i_{1}$ (shown)
which contributes to the last (anomalous) term in (\ref{d22}), together 
with $e_1,f_1,c_1$
(the $B$ contribution vanishes). Inverting the relation
between bare and renormalized disorders yields the 
$\beta$ function $\beta_\Delta=\partial{\Delta} =
\epsilon \Delta + \epsilon \delta^{1}_{div} \Delta
+ \epsilon ( 2 \delta^{2}_{div} \Delta - \delta^{1,1} \Delta )$
where the $1/\epsilon$ terms cancel nicely, the hallmark
of a renormalizable theory ($\delta^{1,1} \Delta$ is the 
counterterm to graph G in Fig.1 and $\partial \equiv -m \partial_m$).
We obtain the $2$-loop FRG equation: 
\begin{eqnarray}
&& \partial{\Delta}(u) =  (\epsilon - 2 \zeta) \Delta(u) 
{+} \zeta u \Delta'(u) {-} \frac{1}{2} \left[(\Delta(u) {-} \Delta(0))^2\right]'' \nonumber \\&&
{+} \frac{1}{2} \left[ (\Delta(u) {-} \Delta(0)) \Delta'(u)^2 \right]'' 
{+} \frac{1}{2} \Delta'(0^+)^2 \Delta''(u) 
\label{rgdisorder}.
\end{eqnarray}
Computing the other needed counterterm, i.e. the renormalized friction $\eta_R =Z^{-1} \eta_0$,
we obtain the dynamical exponent $z  = 2 - \partial \ln Z$.
The $1/\epsilon$ divergences again cancel
yielding the finite result $z = 2 - \Delta''(0^+)
+\Delta''(0^+)^2 + \Delta'''(0^+) \Delta'(0^+) 
(\frac{3}{2} - \ln 2)$.
We stress that (\ref{rgdisorder}) cannot be read at $u=0$ \cite{footnote11}.
Indeed, it (and the cancellation of divergent parts) was established
only for $u \neq 0$. To complete two loop renormalizability 
we checked that  IVF which are $u=0$ quantities 
are also rendered finite by the above counterterms.
We found that the time dependence
in diagrams cancels by subsets as in \cite{footnote6},
i.e. correlations (already rendered finite by the above
procedure) are thus {\em static} for $v=0^+$ at variance with previous works \cite{leschhorn_depinning2}.

For periodic $\Delta(u)$ (CDW depinning \cite{narayan_fisher,unstable})
we find a fixed point of (\ref{rgdisorder}) with 
$\zeta=0$ reading (for a period $1$) $\Delta^*(u) = \frac{\epsilon}{36} + \frac{\epsilon^2}{108}
- (\frac{\epsilon}{6} + \frac{\epsilon^2}{9}) u(1-u)$ ($0<u<1$).
This yields the correlations
$\overline{(u_x-u_0)^2}=A_d \ln|x|$ with 
$A_d =\epsilon/18 + 5 \epsilon^2/108$, the 
RP dynamical exponent $z = 2 - \frac{1}{3} \epsilon - \frac{1}{9} \epsilon^2$
and $\beta=z/2$ from the scaling relation 
\cite{leschhorn_depinning2,narayan_fisher}
$\beta=(z-\zeta)/(2-\zeta)$. $\int_0^1 \Delta^*$ becomes
non zero to two loops, a signature of {\it nonequilibrium
effects}.

Another single FP is found to describe both random field 
and {\em all shorter range disorder}, including RB,
demonstrating the instability of the apparent one loop short range
fixed points. It is determined numerically \cite{us_long}
but $\zeta$ is obtained analytically.  Integrating (\ref{rgdisorder}) over $u>0$
yields $\partial{D} = (\epsilon - 3 \zeta) D - \Delta'(0^+)^3$ where
$D=\int_0^{+\infty} \Delta$ (only assuming $\Delta(+\infty)=0$).
The FP condition then implies \cite{footnote11} (both for RB and RF):
\begin{eqnarray}\label{zeta2loop}
\zeta= \frac{1}{3} \epsilon + \zeta_2 \epsilon^2 = \frac{\epsilon}{3} 
(1 + \frac{\epsilon}{9 \gamma \sqrt{2}})
= \frac{\epsilon}{3} (1 + 0.14331 \epsilon)
\ ,
\end{eqnarray}
where we used that at one loop
$D^* = \sqrt{6}{\epsilon} \gamma \Delta^*(0)^{3/2}$
with $\gamma = \int_0^1 dy \sqrt{y-1-\ln y} = 0.54822$ \cite{chauve_creep}.
This demonstrates a violation of the conjecture 
of \cite{narayan_fisher}. It reconciles theory and
numerical results as shown in Table I where 
the dynamical exponent $z = 2 - \frac{2}{9} \epsilon + \epsilon^2 ( \frac{\zeta_2}{3} 
- \frac{\ln 2}{54} - \frac{5}{108} )
=  2 - \frac{2}{9} \epsilon - 0.04321 \epsilon^2$ as well as
$\beta$ obtained via the scaling relation, $\beta =
1 - \frac{1}{9} \epsilon  - 0.040123 \epsilon^2$,
are also given.

The case of long range elasticity is obtained changing $q^2 + m^2 \to \sqrt{q^2 + m^2}$ in
all propagators, shifting the upper critical dimension to $d_{\rm uc}=2$. It 
yields a renormalizable theory, with $\epsilon=2-d$ and a two loop beta function
\cite{us_long}
obtained by multiplying all ${\cal O}(\Delta^3)$ terms in (\ref{rgdisorder}) by $4 \ln 2$.
This yields $\zeta=\frac{\epsilon}{3}(1 + \frac{4 \ln 2}{9 \gamma \sqrt{2}}\epsilon)
= \frac{\epsilon}{3}(1 + 0.39735 \epsilon)$, i.e. a strong deviation from
$\epsilon/3$ (see Table II), and $z = 1 - \frac{2}{9} \epsilon 
+ \epsilon^2 (  \frac{4 \ln 2}{27 \gamma \sqrt{2}} - \frac{\pi + 20 \ln 2}{108})=
1 - \frac{2}{9} \epsilon - 0.1133 \epsilon^2$. 

We now turn to the {\it statics}, using replicas. In the $T=0$ limit,
the FRG beta function at which we arrive \cite{us_long}:
\begin{eqnarray}
 \partial{R} &=& (\epsilon - 4 \zeta_{\rm eq}) R + \zeta_{\rm eq}  u R' + \frac{1}{2}
R^{\prime \prime 2} - R''(0) R'' \nonumber \\
&& + \frac{1}{2} (R'' - R''(0)) R^{\prime \prime \prime 2}
- \lambda R^{\prime \prime \prime}(0^+)^2 R''
\label{frgr}
\end{eqnarray}
has a new ``anomalous'' term $\propto \lambda$. The other part,
i.e.\ (\ref{frgr}) with $\lambda=0$ (from graphs $a,b$
and repeated one loop counterterm - B graphs cancel in the sum)
could as well be obtained for an analytic $R(u)$, as in
\cite{footnote_larkin}, which by itself would be {\em inconsistent} since the
FP is non-analytic. Ambiguities arise only at two loops
(not at one loop since $R''(0)=R''(0^+)$),  in the 
graphs $e,f$ in Fig.~\ref{fig2} which correct $R(u)$
determining $\lambda$, since some vertices are evaluated at $u=0$.
However we have shown that the theory can be
renormalizable in the usual sense only if:
\begin{eqnarray}
\lambda=1/2 \label{half}
\ .
\end{eqnarray}
Indeed, the form of the repeated one loop counter\-term (i.e. to
G in Fig.1)
$\delta^{1,1} R = [ (R'' - R''(0)) R^{\prime \prime \prime 2}+
(R'' - R''(0))^2 R'''' - R^{\prime \prime \prime}(0^+)^2 R'' ]I^2$
which is {\em non ambiguous} because $\delta^{1} R(u)$ 
is twice differentiable at $u=0$, {\em imposes}
the coefficient of the ambiguous term $e+f$ of $\delta^2 R$
implying (\ref{half}). Interestingly, this value of $\lambda$ 
is also {\em the only one} which prevents the occurrence 
of a further problem in the two loop FRG,
the {\em supercusp} \cite{fractional}. Indeed, e.g.\ in the periodic case, the FP
of (\ref{frgr}) is $R^*(u) = \mbox{const.} - (\frac{\epsilon}{72} 
+ \frac{\epsilon^2}{108}) u^2 (1-u)^2 + 
\frac{\epsilon^2}{432} (2 \lambda-1)  u (1-u)$ and possesses a stronger
singularity than at one loop, since $R^{* \prime}$ is
discontinuous. Thus, unless $\lambda=1/2$, one has  
$\int_0^1 R''= 2 R'(0^+)\not= 0$, i.e.\ 
a violation of potentiality (as naturally occurs above 
in the driven dynamics). The $\lambda=1/2$ theory 
yields $A_d = \frac{\epsilon}{18} + \frac{7 \epsilon^2}{108}$
for one component Bragg glass (and $\int_0^1 \Delta^*$=0
as natural), $\zeta_{\rm eq}=\epsilon/3$ for RF disorder and, via numerics,
$\zeta_{\rm eq}=0.20829804 \epsilon + 0.006858 \epsilon^2$ for RB disorder.
The corresponding extrapolations (Table I) improve the predictions compared to the one loop
result.

Methods aiming at deriving a FRG equation, i.e.\ computing
$\lambda$, beyond (physical) renormalizability or potentiality requirements,
are explored in \cite{us_long}. An alternative exact FRG method, based on multilocal expansion,
also provides \cite{chauve_pld} a procedure
to lift the $u=0$ vertex ambiguities at $T=0$,
and yields (\ref{frgr}) with $\lambda=1/2$ and universal coefficients.
$\lambda=1/2$ is also recovered \cite{chauve_pld} at $T>0$
where it is easy to see how, at large scale where the running
temperature $\tilde{T}_l$ flows to $0$, anomalous terms 
as in (\ref{frgr}) are
generated, e.g.\ from a graph E of 
Fig.~\ref{fig1} (proportional to $\tilde{T}_l R''''(0) R''(u)$)
since the thermal boundary layer analysis at one loop \cite{chauve_creep}
yields $\tilde{T}_l R''''(0) \to R'''(0^+)^2$.

In summary, by finding a way to cope with the difficulties related
to non-analyticity at $T=0$ in the FRG, we obtained the 
exponents characterizing depinning and
statics of pinned elastic systems to next order in 
$\epsilon=4-d$. We predict that similar anomalous
terms arise in other disordered systems where dimensional
reduction fails, e.g. random field spin models.

\unecol


\begin{thebibliography}{999}
\bibitem{footnote6}
noting \cite{chauve_pld} that graphs can be grouped in subsets 
(e.g.\ pairs $a c$, $b d$, $e f$ in Fig.\ \ref{fig2}) which vanish
by shifting 
the endpoint of an internal line within a splitted vertex.
\bibitem{footnote17}
anomalous contributions
(e.g. $\propto \frac{\ln2}{\epsilon}$) are found to cancel.
\bibitem{unstable}
The FP is however unstable in one direction.
\bibitem{footnote11}
This is why the argument of Ref.  \cite{narayan_fisher} ($\zeta=\epsilon/3$ to all orders), fails as it misses
a contribution in
$\int_{-\infty}^{+\infty} du \partial \Delta(u) \neq 2 \partial D$. By contrast at equilibrium,
$\zeta_{\rm eq}=\epsilon/3$ holds to all orders 
if $R^{* \prime}(+\infty) \neq 0$, as it should for RF.
\bibitem{fractional} Fractional powers of disorder then
appear in equilibrium expectation values.
\end{thebibliography}
\end{document}